\documentclass[aps,prl,preprint,superscriptaddress]{revtex4}
\usepackage{float,graphicx}

\begin{document}


\title{The Quantum Geometric Phase between Orthogonal States}


\author{Hon Man Wong}
\email[Email address: ]{hmwong@phy.cuhk.edu.hk}
\affiliation{Department of Physics, The Chinese University of Hong
Kong, Hong Kong SAR, China}

\author{Kai Ming Cheng}
\affiliation{Department of Physics, The Chinese University of Hong
Kong, Hong Kong SAR, China}

\author{M.~-C.~Chu}
\affiliation{Department of Physics, The Chinese University of Hong
Kong, Hong Kong SAR, China}


\date{\today}

\begin{abstract}
We show that the geometric phase between any two states, including
orthogonal states, can be computed and measured using the notion
of projective measurement, and we show that a topological number
can be extracted in the geometric phase change in an infinitesimal
loop near an orthogonal state. Also, the Pancharatnam phase change
during the passage through an orthogonal state is shown to be
either $\pi$ or zero (mod $2\pi$). All the off-diagonal geometric
phases can be obtained from the projective geometric phase
calculated with our generalized connection.
\end{abstract}

\pacs{03.65.Vf, 02.40.Ma}

\maketitle


The existence of the geometric phase was first pointed out by
Berry \cite{berry1984} in adiabatic systems and was later
generalized to non-adiabatic cyclic \cite{aaphase} and non-cyclic
\cite{generalsetting} cases, through the use of the Pancharatnam
connection \cite{generalsetting,pancharatnam} in the latter case.
However, the geometric phase between two orthogonal states is
undefined in the Pancharatnam connection, even though orthogonal
states do contain phase information, which could indeed be
extracted for adiabatic evolution by calculating the off-diagonal
geometric phases \cite{offdiag}.  The latter was generalized
further to non-adiabatic situations by using the $N$-vertex Bargmann
invariant \cite{mukunda2003} and to systems involving
mixed-states \cite{mixedstate}, and it was verified in various experiments
such as microwave cavity \cite{offdiag} and neutron interferometry
\cite{neutron1,neutron2}.

We show in this article that the geometric phase between any two states,
orthogonal or not, can be computed and measured in general using
the concept of projective measurement.
We find that the geometric phase change when a state evolves in an
infinitesimal loop near an orthogonal state is associated with
a topological number, and if the state actually evolves into
an orthogonal state, the phase change is either $\pi$ or zero (mod $2\pi$).
Our generalized connection can be used to calculate all phase
information between two states; in particular,
all the off-diagonal geometric phases can be obtained.

A physical interpretation of the Pancharatnam connection is that
the initial state $\left| {\psi(0)} \right\rangle$ and final state
$\left|\psi (t)\right\rangle$ (with the dynamical phases removed)
are taken to interfere, and the amplitude $\langle \psi(t) |
\psi(0) \rangle $ reflects the phase difference between the
states. Following this idea, we can define another kind of
interference by first projecting the initial and final states to a
certain state $\left| i \right\rangle$, which gives $\left| i
\right\rangle \left\langle {i}
 \mathrel{\left | {\vphantom {i {\psi (0)}}}
 \right. \kern-\nulldelimiterspace}
 {{\psi (0)}} \right\rangle $ and $
\left| i \right\rangle \left\langle {i}
 \mathrel{\left | {\vphantom {i {\psi (t)}}}
 \right. \kern-\nulldelimiterspace}
 {{\psi (t)}} \right\rangle$.
Then they are taken to interfere, giving a projective phase
defined up to an arbitrary phase of $2n\pi$,
\begin{equation}\label{projph}
\varphi _i (0,t) = \arg \left\langle {{\psi (0)}}
 \mathrel{\left | {\vphantom {{\psi (0)} i}}
 \right. \kern-\nulldelimiterspace}
 {i} \right\rangle \left\langle {i}
 \mathrel{\left | {\vphantom {i {\psi (t)}}}
 \right. \kern-\nulldelimiterspace}
 {{\psi (t)}} \right\rangle .
\end{equation}
In particular, when the state $\left|i \right\rangle$ is taken to be
the initial or final state, or some state along the geodesic,
$\varphi_i(0,t)$ reduces to the Pancharatnam phase difference,
$\arg \left\langle {{\psi (0)}}
 \mathrel{\left | {\vphantom {{\psi (0)} {\psi (t)}}}
 \right. \kern-\nulldelimiterspace}
 {{\psi (t)}} \right\rangle.$

An example of this interference is that between the $x$-polarized light
and the $y$-polarized light after directing them through a polarizer in
the direction $\hat x + \hat y $. The
result is physically measurable and is given by Eq.~(\ref{projph}).
The definition in Eq.~(\ref{projph}) is
independent of the choice of gauge of $\left| i \right\rangle$,
since a projection operator, which is gauge invariant, is
used in the definition.

Following the idea in \cite{generalsetting}, the phase can be
evaluated by defining two curves, the projections of which on the
ray space are the shortest geodesics, $\phi_1(s_1)$ and
$\phi_2(s_2)$, $ 0 \le s_k \le 1$, $k=1,2$, which satisfy $ \left|
{\phi _1 (0)} \right\rangle  = \left| {\psi (t)} \right\rangle$, $
\left| {\phi _1 (1)} \right\rangle  = \left| {\phi _2 (0)}
\right\rangle  = \left| i \right\rangle$ and $
 \left| {\phi _2 (1)} \right\rangle  = \left| {\psi (0)} \right\rangle$.
We have therefore
\[
\varphi _i (0,t) = \int_0^1 {A_1 ds_1 }  + \int_0^1 {A_2 ds_2 }
\]
along the two geodesics, where
\[
A_k  = {\mathop{\rm Im}\nolimits} \left\langle {\phi _k (s_k )}
\right|\frac{d}{{ds_k }}\left| {\phi _k (s_k )} \right\rangle,
\]
for $k=1, 2$. Then a closed path $\gamma$ can be defined by the
evolution path of the state and the two geodesics, and the phase in
Eq.~(\ref{projph}) can be evaluated by Stoke's theorem, $\varphi _i
(0,t) = \int_S F$, where $F$ is the two-form $dA$, and the
integral is over the surface $S$ bounded by $\gamma$.
Physically, we first parallel-transport the initial and final states
to $\left| i \right\rangle$ and then compare them to obtain the
phase difference.

Consider the projective measurements at two states, $|i\rangle$
and $|j\rangle$. The projective phases are given by
\begin{equation}
\begin{array}{l}
 \varphi _i (0,t) = \arg \left\langle {0}
 \mathrel{\left | {\vphantom {0 i}}
 \right. \kern-\nulldelimiterspace}
 {i} \right\rangle \left\langle {i}
 \mathrel{\left | {\vphantom {i t}}
 \right. \kern-\nulldelimiterspace}
 {t} \right\rangle , \\
 \varphi _j (0,t) = \arg \left\langle {0}
 \mathrel{\left | {\vphantom {0 j}}
 \right. \kern-\nulldelimiterspace}
 {j} \right\rangle \left\langle {j}
 \mathrel{\left | {\vphantom {j t}}
 \right. \kern-\nulldelimiterspace}
 {t} \right\rangle,  \\
 \end{array}
\end{equation}
where $\left| t \right\rangle  = \left| {\psi (t)} \right\rangle$,
and their difference is,
\[
\varphi _i (0,t) - \varphi _j (0,t) = \arg \left\langle {0}
 \mathrel{\left | {\vphantom {0 i}}
 \right. \kern-\nulldelimiterspace}
 {i} \right\rangle \left\langle {j}
 \mathrel{\left | {\vphantom {j 0}}
 \right. \kern-\nulldelimiterspace}
 {0} \right\rangle  + \arg \left\langle {t}
 \mathrel{\left | {\vphantom {t j}}
 \right. \kern-\nulldelimiterspace}
 {j} \right\rangle \left\langle {i}
 \mathrel{\left | {\vphantom {i t}}
 \right. \kern-\nulldelimiterspace}
 {t} \right\rangle.
\]
We define the gauge transformation, or the transition function as,
\begin{equation}\label{transition_func}
S_{ij} (P) = S_{ji}^{ - 1} (P) = \frac{{\left\langle {j}
 \mathrel{\left | {\vphantom {j P}}
 \right. \kern-\nulldelimiterspace}
 {P} \right\rangle \left\langle {P}
 \mathrel{\left | {\vphantom {P i}}
 \right. \kern-\nulldelimiterspace}
 {i} \right\rangle }}{{\left| {\left\langle {j}
 \mathrel{\left | {\vphantom {j P}}
 \right. \kern-\nulldelimiterspace}
 {P} \right\rangle \left\langle {P}
 \mathrel{\left | {\vphantom {P i}}
 \right. \kern-\nulldelimiterspace}
 {i} \right\rangle } \right|}},
\end{equation}
where $ \left| P \right\rangle $ is a state not orthogonal to
either $\left|i\right\rangle$ or $\left|j\right\rangle$. We
therefore have the transformation,
\[
\exp \left( {i\varphi _i (0,t)} \right) = S_{ij} (0)\exp \left(
{i\varphi _j (0,t)} \right)S_{ji} (t).
\]
$S_{ij}(P)$ depends locally on the state $\left|P\right\rangle$ in
the ray space, as it appears as the projection operator $\left| P
\right\rangle \left\langle P \right| $ in the definition. It
satisfies the properties of the transition function of a
fibre bundle, in regions where $S_{ij}$ is well-defined, i.e., $
S_{ii} (P) = 1$, $ S_{ij} (P)  S_{ji} (P) = 1$ and $ S_{ij} (P)
S_{jk} (P) = S_{ik} (P)$.

We can also calculate the phase change in a segment of the
evolution curve using the definition in Eq.~(\ref{projph}),
\begin{equation} \label{segment_ph}
\varphi _i (t_1,t_2 ) = \arg \left\langle {{t_1 }}
 \mathrel{\left | {\vphantom {{t_1 } i}}
 \right. \kern-\nulldelimiterspace}
 {i} \right\rangle \left\langle {i}
 \mathrel{\left | {\vphantom {i {t_2 }}}
 \right. \kern-\nulldelimiterspace}
 {{t_2 }} \right\rangle .
\end{equation}
This allows us to have associativity not shared by
the Pancharatnam connection,
\begin{eqnarray}
{\varphi _i (0,t)}&=& {\arg \left\langle {0}
 \mathrel{\left | {\vphantom {0 i}}
 \right. \kern-\nulldelimiterspace}
 {i} \right\rangle \left\langle {i}
 \mathrel{\left | {\vphantom {i {t_1 }}}
 \right. \kern-\nulldelimiterspace}
 {{t_1 }} \right\rangle \left\langle {{t_1 }}
 \mathrel{\left | {\vphantom {{t_1 } i}}
 \right. \kern-\nulldelimiterspace}
 {i} \right\rangle \left\langle {i}
 \mathrel{\left | {\vphantom {i t}}
 \right. \kern-\nulldelimiterspace}
 {t} \right\rangle }\nonumber\\
 &=&
{\varphi _i (0,t_1 ) + \varphi _i (t_1 ,t)}
.\label{group_prop_asso}
\end{eqnarray}
When a path goes through a region between $t_1$ and $t_2$ where
the projection on $\left| j \right\rangle$ is defined, but not that of
$\left| i \right\rangle$,
we can easily prove the associative property,
\begin{equation}\label{switch_cover}
e^{i\varphi _i (0,t_3 )}  = e^{i\varphi _i (0,t_1 )} S_{ij} (t_1
)e^{i\varphi _j (t_1 ,t_2 )} S_{ji} (t_2 )e^{i\varphi _i (t_2 ,t_3
)}.
\end{equation}
The covering of $\left|i\right\rangle$ consists of all states not
orthogonal to $\left|i\right\rangle$, and $\varphi_i(0,t)$ is
well-defined when both $\left|\psi(0)\right\rangle$ and
$\left|\psi(t)\right\rangle$ are inside the covering. When the
covering is specified, the structure is a principal coordinate
bundle \cite{gtp}, where the fibre is space of the projective
phase, and the base is the ray space.

For a two-state system, the structure of the ray space is the same
as that of a monopole \cite{gpinp, wuyang}. By the associative
property of Eq.~(\ref{group_prop_asso}), we can write the
projective phase as
\begin{equation}\label{sum_geomph}
\varphi _i (0,t) = \sum\limits_m {\arg \left\langle {{t_m }}
 \mathrel{\left | {\vphantom {{t_m } i}}
 \right. \kern-\nulldelimiterspace}
 {i} \right\rangle \left\langle {i}
 \mathrel{\left | {\vphantom {i {t_{m + 1} }}}
 \right. \kern-\nulldelimiterspace}
 {{t_{m + 1} }} \right\rangle }  = \sum\limits_m {\int_{\Delta S} F } ,
\end{equation}
where $t_{m+1}=t_m + \Delta t$ and $\Delta S$ is the surface
bounded by two geodesics linking $i$ to $\psi(t)$ and
$\psi(t+\Delta t)$ and the evolution path from $\psi(t)$ to
$\psi(t+\Delta t)$. For a two-state system, let
$\left|i\right\rangle=\left|\uparrow\right\rangle$, and we have,
\begin{eqnarray}
   {\int_{\Delta S} F } =  - \frac{1}{2}(1 - \cos \theta )\Delta \phi  ,
\end{eqnarray}
where $\theta$, $\phi$ are the coordinates of the Bloch sphere
and we have used the solid angle formula from \cite{berry1984}. We
can define a vector potential, whose curl is $F$: $
 (A_\phi  )_i  =  - \frac{1}{2}(1 - \cos \theta )$, and $
 (A_\theta  )_i  = 0.$
For $\left| j \right\rangle  = \left|  \downarrow \right\rangle$,
$ (A_\phi  )_j  = \frac{1}{2}(1 + \cos \theta )$,
 $ (A_\theta  )_j  = 0$. The transition function is
$S_{ij} (\theta ,\phi ) = \exp (i\phi )$, where $ \left| {\theta
,\phi } \right\rangle  = \cos {\frac{\theta }{2}} e^{ - i\phi /2}
\left| \uparrow  \right\rangle  + \sin {\frac{\theta}{ 2}}
e^{i\phi /2} \left| \downarrow  \right\rangle$. The results are
formally identical to Wu and Yang's solution to the monopole
problem \cite{wuyang}.

From Eq. (\ref{sum_geomph}), we can write the geometric phase as a
sum of Bargmann invariants,
\begin{eqnarray}
\varphi _i (t,t + \Delta t) &=& \varphi _B (t,i,t + \Delta t)
\nonumber\\
&=& \arg \left\langle {t}
 \mathrel{\left | {\vphantom {t i}}
 \right. \kern-\nulldelimiterspace}
 {i} \right\rangle \left\langle {i}
 \mathrel{\left | {\vphantom {i {t + \Delta t}}}
 \right. \kern-\nulldelimiterspace}
 {{t + \Delta t}} \right\rangle \left\langle {{t + \Delta t}}
 \mathrel{\left | {\vphantom {{t + \Delta t} t}}
 \right. \kern-\nulldelimiterspace}
 {t} \right\rangle.
\end{eqnarray}
The term $ \arg \left\langle {{t + \Delta t}}
 \mathrel{\left | {\vphantom {{t + \Delta t} t}}
 \right. \kern-\nulldelimiterspace}
 {t} \right\rangle$ is responsible for removing the dynamical phase, and
it can be omitted if parallel transport is used. We can construct
a path close to $\left|j\right\rangle$, represented by $ \left|
{\phi (\theta )} \right\rangle  = U\left| j \right\rangle  =
e^{i\delta \hat \lambda (\theta )} \left| j \right\rangle $, where
$\delta$ is a small number and $\hat\lambda$ is a hermitian
operator. The geometric phase change along the path from $\theta
_1$ to $\theta _2$ is,
\begin{eqnarray}\label{logz}
 \varphi _i (\theta _1 ,\theta _2 ) &=& {\mathop{\rm Im}\nolimits} \int_{\theta _1 }^{\theta _2 } {\frac{{\left\langle {\phi }
 \mathrel{\left | {\vphantom {\phi  i}}
 \right. \kern-\nulldelimiterspace}
 {i} \right\rangle \left\langle i \right|\frac{d}{{d\theta }}\left| \phi  \right\rangle }}{{\left| {\left\langle {\phi }
 \mathrel{\left | {\vphantom {\phi  i}}
 \right. \kern-\nulldelimiterspace}
 {i} \right\rangle } \right|^2 }}} d\theta \nonumber \\
  &=& \arg \left. {\left\langle i \right|U\left| j \right\rangle } \right|_{\theta _2 }  - \arg \left. {\left\langle i \right|U\left| j \right\rangle } \right|_{\theta _1 } .
 \end{eqnarray}
On the other hand, the phase $\varphi_j(\theta_1, \theta_2)$ is zero since
the path is infinitesimally close to $\left| j\right\rangle$, and so
\begin{eqnarray}
\varphi_i(\theta_1,
\theta_2)=\arg(S_{ij}(\theta_2)S_{ji}(\theta_1))+2 n \pi .
\end{eqnarray}
Therefore the phases of any two infinitesimal paths with equal
end-points (in the ray space) can differ only by $2 n \pi$. The
phase $\varphi_i(\theta_1, \theta_2)$ is well-defined if and only
if $ z \equiv {\left\langle i \right|U\left| j \right\rangle } $
is not zero at every point on the path. From Eq.~(\ref{logz}) we
can treat the phase $\varphi_i$ as the winding number of $z$ in
its complex plane, and if the path is smoothly deformed with end
points fixed, the value of $\varphi_i$ is not changed, unless the
deformed path of ${U\left| j \right\rangle } $ crosses a zero of
$z$. These zeros naturally divide all paths near
$\left|j\right\rangle$ connecting two fixed points into different
classes corresponding to different phases. This shows that the
phase difference ($2n \pi$) of different classes of paths is
topological.

Although by measurement between two states only the modulo $2\pi$
phase can be measured, the topological part of the phase $2n\pi$
can be observed by accumulating the changes of the phase as a
function of time, $ \varphi_i(t_1, t_2)=\sum {d \varphi_i}/{dt}
\Delta t$, borrowing the idea of continuous measurement of the
Pancharatnam phase in \cite{bhandari2002:1}. We can decompose a
path into segments and measure the phase change as in
Eq.~(\ref{segment_ph}). From this definition, the phase is well
defined if the state does not evolve to a state orthogonal to
$\left|i\right\rangle$ along the path.

If the end-points merge into one ($\theta _2 \to \theta _1$), the
infinitesimal path $U\left|j\right\rangle$ becomes a closed loop,
and the set of geodesics from points on the loop to ${\left| j
\right\rangle }$ defines a two-dimensional manifold $O_j$ that
includes ${\left| j \right\rangle } $.  As $ z={\left\langle i
\right|U\left| j \right\rangle } $ is non-zero, geodesics from
${U\left| j \right\rangle } $ to $\left|i\right\rangle$ are
well-defined \footnote{In Ref.~11 of \cite{generalsetting}, the
geodesic $\phi(s)$ does not go to an orthogonal state as
$\left\langle\phi(0)|\phi(s)\right\rangle$ is real and positive}.
The set of geodesics from ${U\left| j \right\rangle } $ to
$\left|i\right\rangle$ forms a two-dimensional manifold $O_i$.
$O_i$ and $O_j$ form an $S^2$ sphere, and the geometric phase is,
\begin{eqnarray}
 \varphi _i (\theta _1 ,\theta _2 ) &=& 2n\pi = \oint {A_i }=\int_{O_i } {F_i }
 \nonumber\\
  &\approx& \int_{O_i } {F_i }  + \int_{O_j } {F_j }  = \int_{S^2 } F ,
 \end{eqnarray}
where we have used the the approximation that $O_j$ is
infinitesimal.  This is just the two-cell decomposition of the ray space
\cite{gpinqs,mostafazadeh1996}. Therefore, $n$ is the first Chern
number of the constructed $S^2$ sphere, and the
topological properties of the ray space of a system can be
measured by the projective phase near a state orthogonal to the
projection state.

In fact, the topological number $n$ can be measured not only by
infinitesimal loops, but also by finite loops on the constructed $S^2$ sphere.
We have in general,
\[
2n\pi  = \int_{S^2 } F  = \int_{O_i } {F_i }  + \int_{O_j } {F_j }
= \varphi _i - \varphi _j ,
\]
for any {\it finite} loop dividing the $S^2$ sphere into $O_i$ and
$O_j$. This means that the difference in the projective phases
$\varphi _i$ and $\varphi_j$ corresponds to the first Chern number
$n$. The structure is the same as \cite{simon1983} that of
a spin-$S$ system. The closed loop can be arbitrarily deformed,
and $n$ would be invariant as long as the loop does not cross a
zero of $z$.

Furthermore, we can move $\left|i\right\rangle$ continuously so
that $\left|i\right\rangle$ is no longer orthogonal to
$\left|j\right\rangle$, and the first Chern number is still
invariant, as long as every point on the loop is not orthogonal to
either $\left|i\right\rangle$ or $\left|j\right\rangle$.

To illustrate the topological number, we consider a spin-$m$
system, with $S_z=-m, -m+1, ... ,m$.
Let $\left|i\right\rangle=\left|m\right\rangle$ and
$\left|j\right\rangle=\left|-m\right\rangle$. We can evolve the
state around a closed loop near $\left|-m\right\rangle$ from $\left|\theta,
\phi\right\rangle$ to $\left|\theta,\phi+2\pi\right\rangle$, with
$\theta=\pi-\delta$, by a constant magnetic field along the $-\hat z$ direction
with unit magnitude,
such that $H =S_z$. Let $\left|\theta,\phi\right\rangle =d_y
(\theta )\left| m \right\rangle$, where $d_y(\theta)$ is a
rotation by an angle $\theta$ about the $y$-axis. The geometric phase
factor corresponding to $\varphi_i$ is
\begin{eqnarray}
\Phi_i (t) &=& Ne^{i\int {Edt} } \left\langle m \right|e^{ -iS_z t}
d_y (\theta )\left| m \right\rangle\nonumber\\
 &=& Ne^{i( -m + O(\delta ^2 ))t} e^{ - imt} \left\langle {
 m} \right|d_y (\theta )\left| { m} \right\rangle ,
\end{eqnarray}
where $N$ is a normalization constant. Therefore in the limit
$\delta \to 0$, at $t=2\pi$, the total phase change is $-4m\pi$, or
in terms of the change in phase, $ {d\varphi _i (t)}/{d\phi } =  -
2m$, which relates the phase change near an orthogonal state with
the first Chern number $-2m$ in this system \cite{gpinqs,mostafazadeh1996}.
If the curve is not infinitesimally close
to $\left|-m\right\rangle$, we need to compute the quantity
$\varphi_i-\varphi_j$. The corresponding geometric phase factors
are,
\[
\begin{array}{l}
 \Phi _i (t) = Ne^{i\int {Edt} } e^{ - imt} \left\langle m \right|d_y (\beta )\left| m \right\rangle ,  \\
 \Phi _j (t) = Ne^{i\int {Edt} } e^{imt} \left\langle { - m} \right|d_y (\beta )\left| m \right\rangle . \\
 \end{array}
\]
Knowing that $\left\langle { - m} \right|d_y (\beta )\left| m
\right\rangle \neq 0$ if $\beta \neq 0$ or $\pi$, the dynamical
phases cancel, and at $t=2\pi$, we have
\[
\varphi_i-\varphi_j=-2m (2\pi)=-4m\pi ,
\]
which is the same as for the infinitesimal loop.

It is well known that there is a $\pi$ phase jump in the
Pancharatnam phase in a two-state system when the wavefunction
passes through a state orthogonal to the initial state $\left| i
\right\rangle = \left| \uparrow \right\rangle$
\cite{bhandari1991:1,yuen2003}. This can be seen easily using the
projective phase.  As the Pancharatnam phase is undefined at the
orthogonal state, a change of the covering is needed. When the
state evolves near $\left| \downarrow \right\rangle$, the global
gauge is switched to the covering of $\left| j \right\rangle$, and
the phase factor is
\begin{eqnarray}
e^ {i\varphi _i (0,t_2 )}  &=& e^{i\varphi _i (0,t_1 )}S_{ij} (t_1
)e^{i\varphi _j (t_1 ,t_2 )} S_{ji} (t_2 ).\label{phase_i}
\end{eqnarray}
If the path from $t_1$ to $t_2$ is taken to be infinitesimal,
the contribution $\varphi_j(t_1,t_2)$ can be ignored, and the
phase change is simply
\begin{eqnarray}
{S_{ij} (t_1 )S_{ji} (t_2 )}= {\exp \left( {i\varphi _B (t_1
,i,t_2 ,j)} \right)},\label{two_sij}
\end{eqnarray}
where $\varphi_B(a,b,c,d)$ is the four-point Bargmann invariant
\[
\varphi _B (a,b,c,d) \equiv \arg \left\langle {a}
 \mathrel{\left | {\vphantom {a b}}
 \right. \kern-\nulldelimiterspace}
 {b} \right\rangle \left\langle {b}
 \mathrel{\left | {\vphantom {b c}}
 \right. \kern-\nulldelimiterspace}
 {c} \right\rangle \left\langle {c}
 \mathrel{\left | {\vphantom {c d}}
 \right. \kern-\nulldelimiterspace}
 {d} \right\rangle \left\langle {d}
 \mathrel{\left | {\vphantom {d a}}
 \right. \kern-\nulldelimiterspace}
 {a} \right\rangle ,
\]
which is known to be equal to the negative of the geometric phase of the area
enclosed by the four geodesics linking the four states \cite{bargmann1993} .

For a two-state system, we have to choose two coverings to cover
the entire Bloch sphere.  Assuming that the path $\psi(t)$ is
smooth at the orthogonal state, the four geodesics form a great
circle linking $\left|i\right\rangle$ and $\left|j\right\rangle$
on the Bloch sphere. The solid angle $\Omega$ encircled is $2
\pi$, and the corresponding geometric phase is $-\Omega/2=-\pi$.
As a result, Eq.~(\ref{phase_i}) becomes
\begin{equation}\label{eq_pijump}
\varphi _i (0,t_2 ) = \varphi _i (0,t_1 ) - \pi ,
\end{equation}
showing how a sudden $\pi$-jump arises.

In fact, we can give the phase change for a general system with
topology different from $S^2$. If a wavefunction
$\left|\psi(t)\right\rangle$, with
$\left|\psi(0)\right\rangle$=$\left|i\right\rangle$, passes
through an orthogonal state $\left|j\right\rangle$ at time $t_0$,
and $\left|\psi(t)\right\rangle$ evolves according to a continuous
hamiltonian $H(t)$, then near time $t_0$, the state is of the
form,
\[
\left| {\psi (t_0  \pm \delta t)} \right\rangle  =  \left| j
\right\rangle  \pm \left| {\phi _1 } \right\rangle \delta t +
\left| {\phi _2 } \right\rangle \delta t^2  \pm ...  \ .
\]
If the first non-vanishing order is of $\delta t ^p$, then,
\begin{eqnarray}
 \varphi _i (t_0  - \delta t,t_0  + \delta t)
  &=& \arg S_{ij} (t_0  - \delta t) e^{-i \varphi _j} S_{ji} (t_0 + \delta t)\nonumber \\
  &\approx& \arg ( - 1)^p ,
\end{eqnarray}
where we have taken $\langle j \left| {\psi (t_0  \pm \delta t)}
\right\rangle \approx 1$ and $\varphi _j \approx 0$. Therefore the
phase change (mod $2\pi$) is $0$ or $\pi$ passing through an
orthogonal state.

The projective phase defined in Eq.~(\ref{projph}) can be used to
obtain the off-diagonal geometric phases \cite{offdiag}. When a
system evolves adiabatically, and an eigenstate
$\left|\psi_n(s_1)\right\rangle$ evolves to
$\left|\psi_n(s_2)\right\rangle$ which is orthogonal to
$\left|\psi_n(s_1)\right\rangle$,
the off-diagonal geometric phases are defined as
\begin{equation}\label{gammajk}
\gamma _{jk}  = \sigma _{jk}  + \sigma _{kj},
\end{equation}
with $ \sigma _{jk}  = \arg \left\langle {{\psi _j (s_1 )}}
 \mathrel{\left | {\vphantom {{\psi _j (s_1 )} {\psi _k (s_2 )}}}
 \right. \kern-\nulldelimiterspace}
 {{\psi _k (s_2 )}} \right\rangle $ and
 $ \sigma _{kj}  = \arg \left\langle {{\psi _k (s_1 )}}
 \mathrel{\left | {\vphantom {{\psi _k (s_1 )} {\psi _j (s_2 )}}}
 \right. \kern-\nulldelimiterspace}
 {{\psi _j (s_2 )}} \right\rangle$.
For more states, more $\sigma$'s can be
defined, and the independent combinations of $\sigma$'s contain
all the phase information of the system.

\begin{turnpage}
\begin{figure}
\includegraphics[height=10cm]{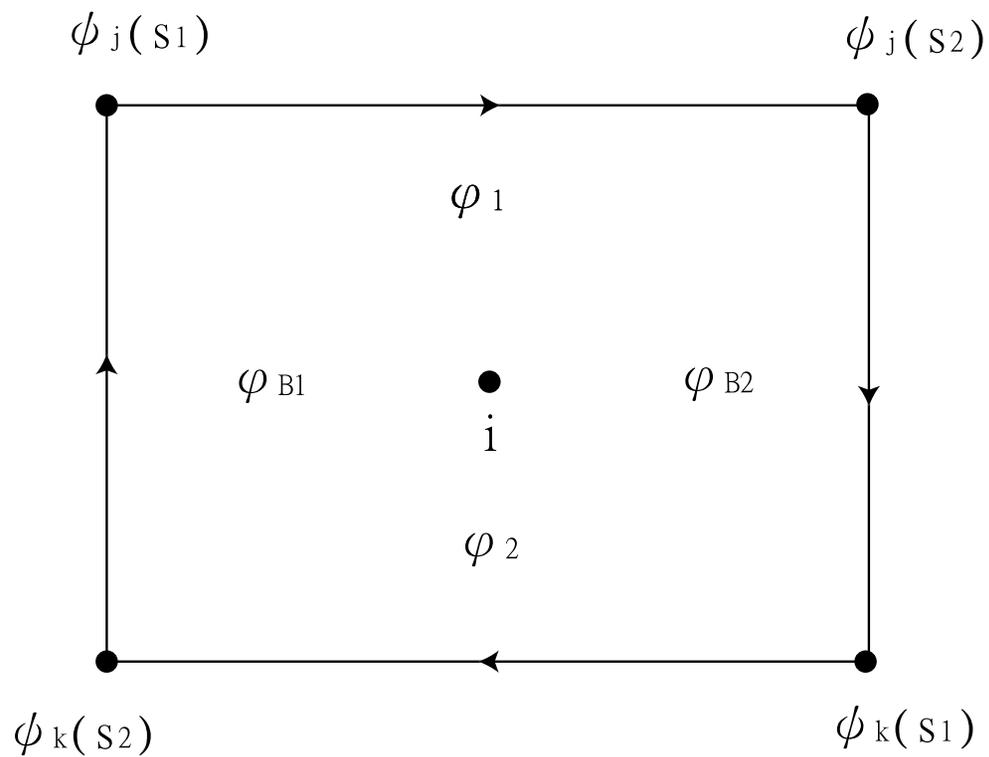}
\caption{The whole area in the figure corresponds to the off-diagonal
geometric phase $\gamma_{jk}$ in Eq.~(\ref{gammajk}). It can be
divided into four parts, each corresponding to a phase in
Eq.~(\ref{off-diag2}), where $\varphi_{B1}=\varphi _B (\psi _j (s_1
),\psi _k (s_2 ),i)$ and $\varphi_{B2}=\varphi _B (\psi _k (s_1
),\psi _j (s_2 ),i)$.\label{offdiag1}}
\end{figure}
\end{turnpage}

When a state evolves to its orthogonal state, we can find a state
$ \left| i \right\rangle$ which is not orthogonal to
$\psi_j(s_1)$, $\psi_j(s_2)$, $\psi_k(s_1)$ or $\psi_k(s_2)$. The
projective phases are,
\begin{equation}\label{phi1phi2}
\begin{array}{l}
 \varphi 1 = \varphi _i (\psi _j (s_1 ),\psi _j (s_2 ))
= \arg \left\langle {{\psi _j (s_1 )}}
 \mathrel{\left | {\vphantom {{\psi _j (s_1 )} i}}
 \right. \kern-\nulldelimiterspace}
 {i} \right\rangle \left\langle {i}
 \mathrel{\left | {\vphantom {i {\psi _j (s_2 )}}}
 \right. \kern-\nulldelimiterspace}
 {{\psi _j (s_2 )}} \right\rangle  ,\\
 \varphi 2 = \varphi _i (\psi _k (s_1 ),\psi _k (s_2 ))
= \arg \left\langle {{\psi _k (s_1 )}}
 \mathrel{\left | {\vphantom {{\psi _k (s_1 )} i}}
 \right. \kern-\nulldelimiterspace}
 {i} \right\rangle \left\langle {i}
 \mathrel{\left | {\vphantom {i {\psi _k (s_2 )}}}
 \right. \kern-\nulldelimiterspace}
 {{\psi _k (s_2 )}} \right\rangle . \\
 \end{array}
\end{equation}
Then the off-diagonal geometric phase is given by,
\begin{eqnarray}
\gamma _{jk}  = &&\varphi _B (\psi _j (s_1 ),\psi _k (s_2 ),i) +
\varphi _B (\psi _k (s_1 ),\psi _j (s_2 ),i)\nonumber\\
 &&+\varphi 1 + \varphi 2.\label{off-diag2}
\end{eqnarray}
As the Bargmann invariants are defined in the ray space, they can
be obtained geometrically, and so can the off-diagonal geometric
phases, using the projective phases $\varphi 1 $ and $\varphi 2$
(see Fig.~\ref{offdiag1}).  Similarly, all the off-diagonal
geometric phases with more states can be obtained from $n$
projective phases. This means that for an $n$-state system, all
$n^2-n+1$ phase relations \cite{offdiag}, including diagonal
(Berry phase) and off-diagonal phases, could be obtained from $n$
projective phases. The projective phase defined here is simple; it
directly gives the phase relation between the initial and final
states, and it does not require knowledge of the hamiltonian or
eigenstates.

The projective phase could be measured by modifying the neutron
interferometry experiment to measure off-diagonal geometric phases
\cite{neutron1}. One of the split paths of the neutron is evolved
by a magnetic field ${\bf B}$ perpendicular to the spin, and the
other is phase shifted by $e^{i\chi}$; then they are brought
together and projected to
$P(\left|i\right\rangle)=\left|i\right\rangle\left\langle
i\right|$. The intensity after interference is
\[
I=\left|e^{i\chi} P(\left|i\right\rangle)\left|\psi(0)
\right\rangle + P(\left|i\right\rangle) U(t)\left|\psi(0)
\right\rangle\right|^2
\]
where $\chi$ is the phase shift by the phase shifter. The
projective phase $\varphi_i(0,t)$ in Eq. (\ref{projph}) can be
extracted from the intensity.

In conclusion, we have defined a measurable geometric phase
between two states, which can be orthogonal, by projecting them
into a state which is not orthogonal to either one. It reduces to
the Pancharatnam phase when a particular projection is chosen.
Furthermore, we show that a topological number is associated with
a closed curve and two projection states. This is
measurable by dividing the loop into small but finite segments and
adding their phase changes together. Also, we used the global
gauge transformation to show that when a state evolves through an
orthogonal state under a continuous Hamiltonian, the phase jump
(mod $2\pi$) can be $\pi$ or $0$ only. Finally, we have reduced
all the off-diagonal geometric phases to $n$ projective phases,
thus showing that there are only $n$ phase relations among $n$
states.



%

%



\begin{acknowledgments}
We thank Prof.~Thomas Au and Mr.~Ho-tak Fung for fruitful discussions.
\end{acknowledgments}


\end{document}